\documentclass[preprint,preprintnumbers,amsmath,amssymb]{revtex4}
\usepackage{graphicx}
\usepackage{color}

\def\beq{\begin{equation}}
\def\eeq{\end{equation}}
\def\eeqn{\end{equation}}
\newcommand\iden{\leavevmode\hbox{\small1\normalsize\kern-.33em1}}

\newcommand{\bea} {\begin{eqnarray}}
\newcommand{\eea} {\end{eqnarray}}

\def\tb {t_\beta}
\def\sb  {s_{\beta}}
\def\cb  {c_{\beta}}

\def\lam{\lambda}

\let\jnfont=\rm
\def\NPB#1,{{\jnfont Nucl.\ Phys.\ B }{\bf #1},}
\def\PLB#1,{{\jnfont Phys.\ Lett.\ B }{\bf #1},}
\def\EPJC#1,{{\jnfont Eur.\ Phys.\ Jour.\ C }{\bf #1},}
\def\PRD#1,{{\jnfont Phys.\ Rev.\ D }{\bf #1},}
\def\PRL#1,{{\jnfont Phys.\ Rev.\ Lett.\ }{\bf #1},}
\def\MPLA#1,{{\jnfont Mod.\ Phys.\ Lett.\ A }{\bf #1},}
\def\JPG#1,{{\jnfont J.\ Phys.\ G }{\bf #1},}
\def\CTP#1,{{\jnfont Commun.\ Theor.\ Phys.\ }{\bf #1},}
\def\JHEP#1,{{\jnfont JHEP \ }{\bf #1},}
\def\NPPS#1,{{\jnfont Nucl.\ Phys.\ Proc.\ Suppl.\ }{\bf #1},}
\def\CPC#1,{{\jnfont Computl.\ Phys.\ Commun.\ }{\bf #1},}
\def\CPL#1,{{\jnfont Chin.\ Phys.\ Lett. }{\bf #1},}
\def\AJS#1,{{\jnfont Astrophys.\ J.\ Suppl. }{\bf #1},}
\def\PR#1,{{\jnfont Phys.\ Rept. }{\bf #1},}
\def\AP#1,{{\jnfont Astropart.\ Phys. }{\bf #1},}
\def\EPL#1,{{\jnfont Europhys.\ Lett. }{\bf #1},}
\def\FP#1,{{\jnfont Fortsch.\ Phys. }{\bf #1},}
\def\JCAP#1 {{\jnfont JCAP \ }{\bf #1} }

\begin{document}

\title{A 95 GeV Higgs boson and spontaneous CP-violation at the finite temperature}
\renewcommand{\thefootnote}{\fnsymbol{footnote}}

\author{Jing Gao, Xiao-Fang Han, Jinghong Ma, Lei Wang, Haotian Xu}
 \affiliation{Department of Physics, Yantai University, Yantai
264005, P. R. China}
\renewcommand{\thefootnote}{\arabic{footnote}}

\begin{abstract}
The ATLAS and CMS collaborations reported a diphoton excess in the invariant mass distribution around the 95.4 GeV with a local significance of $3.1\sigma$.
Moreover, there is another $2.3\sigma$ local excess in the $b\bar{b}$ final state at LEP in the same mass region.
A plausible solution is that the Higgs sector is extended to include an additional Higgs boson with a mass of $95.4$ GeV.
We study a complex singlet scalar extension of the two-Higgs-doublet model in which the 95.4 GeV Higgs is from the mixing of three CP-even Higgs fields.
In addition, the extended Higgs potential can achieve spontaneous CP-violation at the finite temperature and restore CP symmetry at the present
temperature of the Universe. We find that the model can simultaneously explain the baryon asymmetry of the Universe, the diphoton and $b\bar{b}$ excesses around the 95.4 GeV
while satisfying various relevant constraints including the experiments of collider and electric dipole moment.
\end{abstract}
\maketitle

\section{Introduction} 
The baryon asymmetry of the Universe (BAU) is a fundamental question in particle physics
and cosmology. By the observation based on the Big-Bang Nucleosynthesis, the BAU is \cite{pdg2020}
\beq
Y_B \equiv \rho_B/s = (8.2 - 9.2) \times 10^{-11},
\eeq  
 where $s$ is the entropy density, and $\rho_B$ is the baryon number density.
The three necessary Sakharov ingredients for generating such an asymmetry dynamically: baryon number violation, 
C and CP violations, and a departure from thermal equilibrium \cite{Sakharov}.
The electroweak baryogenesis (EWBG) \cite{ewbg1,ewbg2} is a promising and
attractive mechanism of explaining the BAU.
In order to achieve the EWBG, the SM need be extended to produce sufficient large CP-violation and a strongly first-order electroweak phase transition (PT),
such as the singlet extension of SM (see for example \cite{bgs-1,bgs-3,bgs-4,bgs-5,Xiao:2015tja,bgs-6,Beniwal:2018hyi,Huang:2018aja,Ghorbani:2017jls,cao,huang,Xie:2020wzn,Ellis:2022lft,Lewicki:2021pgr,Idegawa:2023bkh,Harigaya:2022ptp}) 
and the two-Higgs-doublet model (2HDM) (see for example \cite{bg2h-1,bg2h-2,bg2h-3,Kanemura:2004ch,Basler:2017uxn,Abe:2013qla,bg2h-5,bg2h-4,bg2h-6,bg2h-7,bg2h-8,bg2h-9,bg2h-11,Basler:2020nrq,Ge:2020mcl,bg2h-13,bg2h-12,2111.13079,2207.00060,Goncalves:2023svb,Li:2024mts}).

The explicit CP-violation interactions in the lagrangian required by the EWBG can be severely constrained by the negative
results in the electric dipole moment (EDM) searches for electrons \cite{edm-e}. 
The spontaneous CP-violation at finite temperature offers a valid mechanism for solving the problem.
In the mechanism, no explicit CP-violating interactions are added to the lagrangian. Because of thermal corrections to the effective potential,
the vacuum structure of the theory changes at finite temperature, and can dynamically induce sufficient large CP-violation to obtain the observed BAU via electroweak PT. 
However, as the temperature decreases, the observed electroweak vacuum is obtained and the CP symmetry is restored at the present temperature. Thus, 
there are no new CP-violating interactions beyond the CKM mechanism at the present temperature, and no new corrections to the EDM. 
As a result, the stringent EDM experimental constraints will be naturally avoided.
The novel mechanism has been realized in the singlet scalar extension of the SM \cite{cao,huang} in which one need introduce a high dimension effective
operator, the singlet pseudoscalar extension of 2HDM \cite{Huber:2022ndk,Liu:2023sey}, and the complex singlet scalar extension of 2HDM respecting a discrete dark CP symmetry
in which the dark matter is simultaneously explained \cite{Ma:2023kai}. 

Based on full Run 2 data set, the CMS collaboration published the latest result of searching for low-mass diphoton,
and confirmed a previously released excess at $m_{\gamma\gamma}$ = 95.4 GeV with
a local significance of $2.9\sigma$ \cite{cms95}. The ATLAS collaboration
 also presented the result of searching for
diphoton signals in the mass range from 66 GeV to 110 GeV based on their full Run 2 dataset \cite{atlas95}. 
The ATLAS search found a diphoton excess with a local significance of $1.7\sigma$
at the same mass value as the one that was previously reported by CMS. Neglecting possible correlations, the
combined signal strength corresponds to a $3.1\sigma$ local excess \cite{Biekotter:2023oen},
\beq
\mu_{\gamma\gamma}^{exp}=\mu_{\gamma\gamma}^{ATLAS+CMS}=\frac{\sigma(pp\to\phi\to \gamma\gamma)}{\sigma_{SM}(pp\to h^{SM}_{95.4}\to \gamma\gamma)}=0.24^{+0.09}_{-0.08},
\eeq
where $\phi$ is a non-SM scalar with a mass of 95.4 GeV, and $h^{SM}_{95.4}$ is a hypothetical SM-like Higgs with the same mass.
Interestingly, there is a longstanding $2.3\sigma$ local excess in the $e^+ e^- \to Z(\phi\to b\bar{b})$ searches at LEP in the same mass region \cite{LEPWorkingGroupforHiggsbosonsearches:2003ing},
\beq
\mu_{b\bar{b}}^{exp}=0.117\pm 0.057.
\eeq

There are numerous discussions on the excesses in the new physics models, for example Refs. \cite{Biekotter:2023oen,Cao:2016uwt,Biekotter:2019kde,Heinemeyer:2021msz,Biekotter:2022jyr,
Aguilar-Saavedra:2023tql,Choi:2019yrv,Ma:2020mjz,Li:2022etb,Ellwanger:2023zjc,Dev:2023kzu,Bonilla:2023wok,Liu:2024cbr,Li:2023kbf,Abbas:2023bmm,
Ahriche:2023hho,Cao:2023gkc,Maniatis:2023aww,Belyaev:2023xnv,Azevedo:2023zkg,Bhatia:2022ugu,Chen:2023bqr,Ahriche:2023wkj,Arcadi:2023smv,
Borah:2023hqw,Banik:2023vxa,Dutta:2023cig,Bhattacharya:2023lmu,Ashanujjaman:2023etj,Escribano:2023hxj,Biekotter:2023jld,
Banik:2023ecr,Coloretti:2023wng,Ahriche:2022aoj,Benbrik:2022azi,Biekotter:2022abc,Biekotter:2021qbc,Abdelalim:2020xfk,Biekotter:2020cjs,
Aguilar-Saavedra:2020wrj,Cao:2019ofo,Kundu:2019nqo,Cao:2024axg,Kalinowski:2024uxe,Ellwanger:2024vvs,Benbrik:2024ptw,Lian:2024smg,Yang:2024fol}.
 In this paper, we propose a complex singlet scalar extension of 2HDM (2HDMS) to accommodate the diphoton and $b\bar{b}$ excesses around 95.4 GeV.
Meanwhile, we discuss the spontaneous CP-violation at the finite temperature and possibility of explanation on the BAU. 

The structure of this paper is outlined as follows. In Section II, we introduce some characteristic features
of the 2HDMS. In Section III and IV, we discuss relevant theoretical and
experimental constraints, and explain the diphoton and $b\bar{b}$ excesses around the 95.4 GeV. 
In Section V and VI, we discuss the PTs of the Universe and the possibility of explaining the BAU, respectively.
Finally, we draw our conclusion in Section VII.

\section{The model} 
The SM is augmented with a second Higgs doublet $\Phi_2$ and a complex singlet $S$,
\bea
&&\Phi_1=\left(\begin{array}{c} \phi_1^+ \\
\frac{(v_1+\rho_1+i\eta_1)}{\sqrt{2}}\,
\end{array}\right)\,, 
\Phi_2=\left(\begin{array}{c} \phi_2^+ \\
\frac{(v_2+\rho_2+i\eta_2)}{\sqrt{2}}\,
\end{array}\right),
S=\frac{(\rho_s+v_s+i\eta_s)}{\sqrt{2}},
\eea
where $v_1$, $v_2$ and $v_s$ are the real vacuum expectation values (VEVs) acquired by the fields $\Phi_1, \Phi_2$ and $S$, respectively.
We define the ratio of the two VEVs as $\tan\beta \equiv v_2 /v_1$.

The scalar potential is given by
\begin{eqnarray} \label{V2HDM} &&\mathrm{V} = m_{11}^2
(\Phi_1^{\dagger} \Phi_1) + m_{22}^2 (\Phi_2^{\dagger}\Phi_2) + \frac{\lambda_1}{2}  (\Phi_1^{\dagger} \Phi_1)^2 +
\frac{\lambda_2}{2} (\Phi_2^{\dagger} \Phi_2)^2 + \lambda_3
(\Phi_1^{\dagger} \Phi_1)(\Phi_2^{\dagger} \Phi_2) \nonumber \\
&& + \lambda_4
(\Phi_1^{\dagger}
\Phi_2)(\Phi_2^{\dagger} \Phi_1)+ \left[\frac{\lambda_5}{2} (\Phi_1^{\dagger} \Phi_2)^2  +  \lambda_6 (\Phi_1^{\dagger} \Phi_1) (\Phi_1^{\dagger} \Phi_2) + \lambda_7 (\Phi_2^{\dagger} \Phi_2) (\Phi_1^{\dagger} \Phi_2)  +\rm
h.c.\right]\nonumber\\
&&+m^2_{S}SS^* + \left[\frac{m^{\prime 2}_{S}}{2}SS+\rm h.c. \right] +  \left[ \frac{\mu_s}{6}S^3 + \frac{\mu^{\prime}_s}{2}SSS^*  + \rm h.c.\right]\nonumber\\
&&+ \left[\frac{\lambda^{\prime\prime}_1}{24}S^4 + \frac{\lambda^{\prime\prime}_2}{6}S^2SS^*+\rm h.c. \right] +\frac{\lambda^{\prime\prime}_3}{4}(SS^*)^2 \nonumber\\
&&+SS^*\left[\lambda^{\prime}_1\Phi_1^{\dagger} \Phi_1+\lambda^{\prime}_2\Phi_2^{\dagger} \Phi_2 \right]
+\left[S^2 (\lambda^{\prime}_4\Phi_1^{\dagger} \Phi_1+\lambda^{\prime}_5\Phi_2^{\dagger} \Phi_2) +\rm h.c. \right]\nonumber\\
&&+ \left[(\lambda^{\prime}_6 SS^* + \lambda^{\prime}_7 SS + \lambda^{\prime}_8 S^*S^*) \Phi_2^{\dagger} \Phi_1 + \rm h.c. \right]\nonumber\\
&&+ \left[-m_{12}^2 \Phi_2^{\dagger} \Phi_1 + \frac{\mu}{2} (S-S^*) \Phi_1^{\dagger} \Phi_2  + \rm h.c. \right].
\end{eqnarray}
We focus on the CP-conservation case in which all the coupling coefficients and mass are real.
For simplicity, we take $\lambda_6=\lambda_7=\lambda^\prime_6=\lambda^\prime_7=\lambda^\prime_8=0$ and $\mu_s=\mu^{\prime}_s$ in the following discussions.
A softly-broken $Z_2$ symmetry under which $\Phi_1\to \Phi_1, S\to S$ and $\Phi_2\to - \Phi_2$, can be imposed to achieve $\lambda_6=\lambda_7=\lambda^\prime_6=\lambda^\prime_7=\lambda^\prime_8=0$.


The minimization conditions of the scalar
potential give
\bea
\quad m_{11}^2 &=& m_{12}^2 \tb - \frac{v^2}{2} \left( \lambda_1 \cb^2 + \lambda_{345}\sb^2 \right)- \frac{v_{s}^2}{2}\lambda^{\prime}_1- v_{s}^2\lambda^{\prime}_4\,,\nonumber\\
\quad m_{22}^2 &=& m_{12}^2 / \tb -  \frac{v^2}{2} \left( \lambda_2 \sb^2 + \lambda_{345}\cb^2 \right)-\frac{v_{s}^2}{2}\lambda^{\prime}_2-  v_{s}^2\lambda^{\prime}_5,\nonumber\\
\quad m_{S}^2 &=& -m^{\prime 2}_{S}-\frac{v_{s}^2}{12}\lambda^{\prime\prime}_1-\frac{v_{s}^2}{3}\lambda^{\prime\prime}_2-\frac{v_{s}^2}{4}v_{s}^2\lambda^{\prime\prime}_3-\sqrt{2}v_{s}\mu_s \nonumber\\
\quad && - \frac{1}{2} v^2 ( \lambda^{\prime}_1+2\lambda^{\prime}_4)\cb^2 - \frac{1}{2} v^2(\lambda^{\prime}_2+2\lambda^{\prime}_5)\sb^2,
\label{min_cond}
\eea
where $\tb\equiv \tan\beta$, $\sb\equiv \sin\beta$, $\cb \equiv \cos\beta$,
and $\lambda_{345} = \lambda_3+\lambda_4+\lambda_5$.

After spontaneous symmetry breaking, the mass eigenstates are obtained from the original fields by the rotation matrices,
\begin{eqnarray}
\left(\begin{array}{c}h_1,h_2,h_3 \end{array}\right) =   \left(\begin{array}{c} \rho_1,  \rho_2,  \rho_s\end{array} \right) R^T, \\
\left(\begin{array}{c}A,~X,~G \end{array}\right) =   \left(\begin{array}{c} \eta_1,  \eta_2,  \eta_s\end{array} \right) R^T_A, \\
\left(\begin{array}{c}G^{\pm} \\ H^{\pm} \end{array}\right) =  \left(\begin{array}{cc}\cos\beta & \sin\beta \\ -\sin\beta & \cos\beta \end{array}\right)  \left(\begin{array}{c} \phi^{\pm}_1 \\ \phi^{\pm}_2 \end{array}\right),
\end{eqnarray}
with 
\beq
	R= \left( \begin{array}{*{20}{c}}
			c_{1}   c_{2} & s_{1}  c_{2}  & s_{2}\\
			  -  s_{1}  c_{3}-c_{1}   s_{2}   s_{3}  &  c_{1}  c_{3}-s_{1}  s_{2}  s_{3}   & c_{2}  s_{3}  \\
			 s_{ 1}  s_{3}-c_{1}  s_{2}  c_{3} &- s_{1}  s_{2}c_{3}  -c_{1}  s_{3}  & c_{2}  c_{3}  
	\end{array} \right),R^A= \left( \begin{array}{*{20}{c}}
			-s_\beta  c_{4} & c_\beta  c_{4} & s_{4}\\
			    s_\beta s_{4} & -c_\beta  s_{4}  & c_{4}  \\
			 c_\beta    & s_\beta  & 0  
	\end{array} \right).
\eeq
The shorthand notations $s_{1}\equiv\sin\alpha_1$, $s_{2}\equiv\sin\alpha_2$, $s_{3}\equiv\sin\alpha_3$, and $s_{4}\equiv\sin\alpha_4$.
The $G^0$ and $G^\pm$ are Goldstones, and they are absorbed by gauge bosons $Z$ and $W^\pm$. 
The remaining physical states includes three CP-even states $h_{1,2,3}$,
 two pseudoscalars $A$ and $X$, and a pair of charged scalar $H^{\pm}$.

Taking the scalar masses and mixing angles as the input parameters, one can express the coupling
constants in the Higgs potential as
\begin{eqnarray}\label{poten-cba}
&&\mu =\frac{\sqrt{2}c_{4} s_{4}(m_{X}^2-m_{A}^2)}{v}, \nonumber\\
&& \lambda_1  = \frac{ m_{h1}^2 R_{11}^2+m_{h2}^2 R_{21}^2 +m_{h3}^2 R_{31}^2  - m_{12}^2 t_\beta}{v^2 c_\beta^2},  \nonumber \\  
&& \lambda_2 = \frac{  m_{h1}^2 R_{12}^2 +m_{h2}^2  R_{22}^2+m_{h3}^2 R_{32}^2- m_{12}^2 t_\beta^{-1}}{v^2 s_\beta^2},  \nonumber \\  
&&\lambda_3 =  \frac{m_{h1}^2 R_{11}R_{12}+m_{h2}^2 R_{21}R_{22}+m_{h3}^2 R_{31}R_{32}+ 2 m_{H^{\pm}}^2 s_\beta c_\beta - m_{12}^2}{v^2 s_\beta c_\beta },  \nonumber \\ 
&&\lambda_4 =  \frac{(m_A^2c_4^2 + m_X^2 s_4^2 -2m_{H^{\pm}}^2 ) s_\beta c_\beta + m_{12}^2}{v^2 s_\beta c_\beta },   \nonumber \\    
&&\lambda_5=  \frac{ - (m_A^2c_4^2 + m_X^2 s_4^2) s_\beta c_\beta  + m_{12}^2}{ v^2 s_\beta c_\beta },\nonumber \\ 
&&\lambda^{\prime}_4=-\frac{3 c_{4}^2 m_{X}^2+6{m^{\prime 2}_{S}}+3m_{A}^2 s_{4}^2+6\lambda^{\prime}_5 {s_\beta^2} v^2+3\sqrt{2}{\mu_s}v_s+(\lambda^{\prime\prime}_1+\lambda^{\prime\prime}_2)v_{s}^2}
 {{6v^2 c_\beta^2}} ,  \nonumber\\
&&\lambda^{\prime}_1=-2\lambda^{\prime}_4 + \frac{(m_{h1}^2 R_{11} R_{13}+m_{h2}^2 R_{21} R_{23}+m_{h3}^2 R_{31} R_{33})}{v v_s c_\beta} ,  \nonumber\\
&&\lambda^{\prime}_2=-2\lambda^{\prime}_5 + \frac{(m_{h1}^2 R_{12} R_{13}+m_{h2}^2 R_{22} R_{23}+m_{h3}^2 R_{32} R_{33})}{v v_s s_\beta},  \nonumber \\
&&\lambda^{\prime\prime}_3=\frac{6m_{h1}^2 R_{13}^2+6m_{h2}^2 R_{23}^2+6m_{h3}^2 R_{33}^2-v_s(6 \sqrt{2}\mu_s+\lambda^{\prime\prime}_1 v_s+4\lambda^{\prime\prime}_2 v_s)}{3v_{s}^2}.  \nonumber\\
\label{eq:lambdas}
\end{eqnarray}

The general Yukawa interactions are written as
 \bea
&&- {\cal L} =Y_{u2}\,\overline{Q}_L \, \tilde{{ \Phi}}_2 \,u_R
+\,Y_{d2}\,
\overline{Q}_L\,{\Phi}_2 \, d_R\, + \, Y_{\ell 2}\,\overline{L}_L \, {\Phi}_2\,e_R \,\nonumber\\
&&+Y_{u1}\,\overline{Q}_L \, \tilde{{ \Phi}}_1 \,u_R
+\,Y_{d1}\,
\overline{Q}_L\,{\Phi}_1 \, d_R\, + \, Y_{\ell 1}\,\overline{L}_L \, {\Phi}_1\,e_R+\, \mbox{h.c.},
\eea where
$Q_L^T=(u_L\,,d_L)$, $L_L^T=(\nu_L\,,l_L)$,
$\widetilde\Phi_{1,2}=i\tau_2 \Phi_{1,2}^*$, and $Y_{u1,2}$,
$Y_{d1,2}$ and $Y_{\ell 1,2}$ are $3 \times 3$ matrices in family
space. 
 We assume the Yukawa coupling matrices to be aligned to avoid the tree-level flavour changing neutral current \cite{aligned2h,Wang:2013sha},
 \bea
 &&(Y_{u1})_{ii}=\frac{\sqrt{2}m_{ui}}{v}\rho_{1u}, ~~~ (Y_{u2})_{ii}=\frac{\sqrt{2}m_{ui}}{v}\rho_{2u},\nonumber\\
&&(Y_{\ell 1})_{ii}=\frac{\sqrt{2}m_{\ell i}}{v}\rho_{1\ell}, ~~~~ (Y_{\ell 2})_{ii}=\frac{\sqrt{2}m_{\ell i}}{v}\rho_{2\ell},\nonumber\\
&&(X_{d1})_{ii}=\frac{\sqrt{2}m_{di}}{v}\rho_{1d},  ~~~ (X_{d2})_{ii}=\frac{\sqrt{2}m_{di}}{v}\rho_{2d},
\eea
where all the off-diagonal elements are zero, and $\rho_{1f}=(c_\beta-s_\beta \kappa_f)$ and $\rho_{2f}=(s_\beta+c_\beta \kappa_f)$ with $f=u,d,\ell$. 
$\kappa_f$ are new free parameters, which determine the Yukawa coupling matrices of up-type quark, down-type quark, and lepton.
$X_{d1,2}=V_{dL}^\dagger Y_{d1,2} V_{dR}$ with $V_{dL}\equiv V_{CKM}$, where $V_{dL,R}$ are unitary matrices which transform the interaction eigenstates to the mass
eigenstates for the left-handed and right-handed down-type quark fields. 
The Yukawa interactions will explicitly break the $Z_2$ symmetry mentioned above.

The couplings of the neutral Higgs bosons with respect to the SM are given by
\bea\label{hffcoupling} &&
y^{h_1}_V=c_2 c_{\beta 1},~y_{f}^{h_1}=c_2 \left(c_{\beta 1}- s_{\beta 1}\kappa_f\right), \nonumber\\
&&y^{h_2}_V \simeq |s_2| s_{\beta 13}+\frac{c_2^2}{2}c_3 s_{\beta 1},~y^{h_2}_f \simeq |s_2| \left(s_{\beta 13}+c_{\beta 13}\kappa_f\right)+\frac{c_2^2}{2}c_3\left(s_{\beta 1} + c_{\beta 1}\kappa_f\right), \nonumber\\
&&y^{h_3}_V \simeq |s_2| c_{\beta 13}-\frac{c_2^2}{2}c_3 s_{\beta 1},~y^{h_3}_f \simeq |s_2| \left(c_{\beta 13}-s_{\beta 13}\kappa_f\right)-\frac{c_2^2}{2}c_3\left(s_{\beta 1} + c_{\beta 1}\kappa_f\right), \nonumber\\
&&y^{A}_V=0,~y_{A}^{f}=-i\kappa_f~{\rm (for}~u)c_4,~y_{f}^{A}=i \kappa_fc_4~{\rm (for}~d,~\ell),\nonumber\\ 
&&y^{X}_V=0,~y_{X}^{f}=i\kappa_f~{\rm (for}~u)s_4,~y_{f}^{X}=-i \kappa_fs_4~{\rm (for}~d,~\ell),
\eea 
where $V$ denotes $Z$ or $W$. The shorthand notations $c_{\beta 1}\equiv\cos(\beta-\alpha_1)$, $s_{\beta 1}\equiv\sin(\beta-\alpha_1)$, 
$c_{\beta 13}\equiv\cos(\beta-\alpha_1-sgn(s_2)\alpha_3)$, and $s_{\beta 13}\equiv\sin(\beta-\alpha_1-sgn(s_2)\alpha_3)$.
We take the approximation of $s_2\simeq sgn(s_2)(1-\frac{c_2^2}{2})$ for the expressions in the second and third lines of Eq. (\ref{hffcoupling})

The Yukawa couplings of the charged Higgs are 
\begin{align} \label{yukawa-charge}
 \mathcal{L}_Y & = - \frac{\sqrt{2}}{v}\, H^+\, \Big\{\bar{u}_i \left[\kappa_d\,(V_{CKM})_{ij}~ m_{dj} P_R
 - \kappa_u\,m_{ui}~ (V_{CKM})_{ij} ~P_L\right] d_j + \kappa_\ell\,\bar{\nu} m_\ell P_R \ell
 \Big\}+h.c.,
 \end{align}
where $i,j=1,2,3$ are the index of generation.

\section{Relevant theoretical and experimental constraints}
In our calculations, we consider the following theoretical and experimental constraints:

{\bf (1) The signal data of the 125 GeV Higgs.} 
We take $h_2$ as the observed 125 GeV state, and apply $\textsf{HiggsTools}$ \cite{Bahl:2022igd}to calculate the total
$\chi^2$ of the latest LHC rate measurements of the observed Higgs boson at 125 GeV. $\textsf{HiggsTools}$ incorporates the codes
$\textsf{HiggsSignals}$ \cite{Bahl:2022igd,Bechtle:2013xfa} and $\textsf{HiggsBounds}$ \cite{Bechtle:2020pkv,Bechtle:2008jh}. 
We pay particular attention to the surviving samples with $\chi^2-\chi^2_{SM}<$ 6.18, where $\chi^2_{SM}$ denotes the $\chi^2$ value of the SM. 
These samples are favoured compared to the SM
fit result within $2\sigma$ confidence level (assuming two-dimensional parameter estimations).

{\bf (2) The searches for extra Higgses at the collider and the flavor observables.} 
We employ $\textsf{Higgstools}$ to pick out the samples which meet the 95\% confidence level exclusion limits from searches for 
the additional Higgs at the collider. $\textsf{SuperIso-v4.1}$ \cite{Mahmoudi:2008tp} is used to check the compatibility with $B\to X_s\gamma$ transitions

{\bf (3) Vacuum stability.} 
We impose the conditions that the vacuum should be a global minimum of the potential, and the potential remains positive when the
field values approach infinity, namely boundedness from below. For large values of the fields, the quadratic and cubic terms can be neglected, and the quartic part 
 $V_{4-min}$ in Eq. (\ref{V2HDM}) is written in matrix form in the basis 
$B=\begin{pmatrix} \Phi_1^\dagger \Phi_1 , & \Phi_2^\dagger \Phi_2 , & \rho_S^2 , & \eta_S^2 \end{pmatrix}^T$,
\begin{align}
    V_{4-min}
    &=C^T \frac{1}{2} \underbrace{\begin{pmatrix}
    \lambda_1 & 
    \lambda_3 + \Delta & 
    \lambda_1' + 2\lambda_4' & 
    \lambda_1' - 2\lambda_4' \\
    \lambda_3 + \Delta & 
    \lambda_2 & 
    \lambda_2' + 2\lambda_5' & 
    \lambda_2' - 2\lambda_5' \\
    \lambda_1' + 2\lambda_4' & 
    \lambda_2' + 2\lambda_5' & 
    \frac{\lambda_1'' + 3\lambda_3''+4\lambda_2''}{6} & 
    \frac{-\lambda_1'' + \lambda_3''}{2} \\
    \lambda_1' - 2\lambda_4' & 
    \lambda_2' - 2\lambda_5' & 
    \frac{-\lambda_1'' + \lambda_3''}{2} & 
    \frac{\lambda_1'' + 3\lambda_3''-4\lambda_2''}{6}
    \end{pmatrix}}_{A} C \nonumber \\
    &= \frac{1}{2} C^T A C ,
    \label{bfbeq}
\end{align}
with $\Delta=0$ for $\lambda_4 \geq |\lambda_5|$ and $\Delta=\lambda_4 - |\lambda_5|$ for $\lambda_4 < |\lambda_5|$.
Boundedness from below demands $A$ is a copositive matrix.
Deleting the $i$-th row and
column of $A$, $i=1,2,3,4$, and four symmetric $3\times 3$ matrices are remained.
 The copositivity of the symmetric order 3 matrix $B$ with entries $b_{ij}$, $i,j=1,2,3$ demands \cite{Kannike:2012pe,Dutta:2023cig}
    \begin{align}
        &b_{11} \geq 0 , \quad b_{22} \geq 0 , \quad b_{33} \geq 0,  \nonumber\\ 
        &\Bar{b}_{12} = b_{12} + \sqrt{b_{11}b_{22}} \geq 0,  \nonumber\\
        &\Bar{b}_{13} = b_{13} + \sqrt{b_{11}b_{33}} \geq 0,  \nonumber\\
        &\Bar{b}_{23} = b_{23} + \sqrt{b_{22}b_{33}} \geq 0, \nonumber\\
        &\sqrt{b_{11}b_{22}b_{33}} + b_{12}\sqrt{b_{33}} + b_{13}\sqrt{b_{22}} + b_{23}\sqrt{b_{11}} + \sqrt{2 \Bar{b}_{12} \Bar{b}_{13} \Bar{b}_{23}} \geq 0. \label{bfb5}
    \end{align}

Besides, the copositivity of matrix $A$ requires
 $\det(A) \geq 0  \lor  (\text{adj} A)_{ij} < 0$, for some $i,j$. 
 The adjugate of $A$ is the transpose of the cofactor matrix of $A$, $(\text{adj} A)_{ij} = (-1)^{i+j}M_{ji}$. The $M_{ij}$ is the determinant of 
the submatrix which is obtained by deleting row $i$ and column $j$ of $A$.

{\bf (4) Tree-level perturbative unitarity.} 
We require that the eigenvalues of the $2\to 2$ scalar-scalar scattering matrix are below $8\pi$, 
which is achieved in $\textsf{SPheno-v4.0.5}$ \cite{Porod:2003um} employing $\textsf{SARAH-SPheno files}$ \cite{Staub:2013tta}.

{\bf (5) The oblique parameters.}
The oblique parameters can impose constraints on the
Higgs mass spectrum in the model. For points being in agreement with the experimental observation, it was required
that the values of the $S$ and the $T$ parameter are within the $2\sigma$ ranges, corresponding to
$\chi^2$ = 6.18 for two degrees of freedom.
The fit results are given in Ref. \cite{pdg2020},
\beq
S=0.00\pm 0.07,~~  T=0.05\pm 0.06, 
\eeq
with the correlation coefficients, $\rho_{ST}$ = 0.92.

In the model, the $S$ and $T$ are approximately calculated by \cite{stu1,stu2}
\bea\label{stu-eq}
S&=&\frac{1}{\pi m_Z^2}\left[\sum_{i=1,2,3}\left( F_S(m_Z^2,m_{h_i}^2,m_A^2) (R^A_{11} R_{i1} + R^A_{12} R_{i2})^2 \right. \right.\nonumber\\
&&\left. +F_S(m_Z^2,m_{h_i}^2,m_X^2) (R^A_{21} R_{i1} + R^A_{22} R_{i2})^2  +F_S(m_Z^2,m_Z^2,m_{h_i}^2) (c_\beta R_{i1} +s_\beta R_{i2})^2 \right. \nonumber \\
&&\left.\left.-m_Z^2 F_{S0}(m_Z^2,m_Z^2,m_{h_i}^2) (c_\beta R_{i1} +s_\beta R_{i2})^2 \right. \right)\nonumber \\
&&\left.  - F_S(m_{Z}^2,m_{H^{\pm}}^2,m_{H^{\pm}}^2) -F_S(m_Z^2,m_Z^2,m_{ref}^2)  + m_Z^2 F_{S0}(m_Z^2,m_Z^2,m_{ref}^2)\right], \nonumber\\
T&=&\frac{1}{16\pi m_W^2 s_W^2} \left[\sum_{i=1,2,3}\left( -F_T(m_{h_i}^2,m_{A}^2) (R^A_{11} R_{i1} + R^A_{12} R_{i2})^2 \right. \right.\nonumber\\
&&\left.-F_T(m_{h_i}^2,m_{X}^2) (R^A_{21} R_{i1} + R^A_{22} R_{i2})^2 + F_T(m_{H^{\pm}}^2,m_{h_{i}}^2) (c_\beta R_{i2} -s_\beta R_{i1})^2 \right. \nonumber \\
&&\left.\left.+ 3 F_T(m_{Z}^2,m_{h_{i}}^2) (c_\beta R_{i1} +s_\beta R_{i2})^2 -3 F_T(m_{W}^2,m_{h_{i}}^2) (c_\beta R_{i1} +s_\beta R_{i2})^2\right. \right)\nonumber \\
&&\left.  + F_T(m_{H^{\pm}}^2,m_A^2)(c_\beta R^A_{12} -s_\beta R^A_{11})^2 +  F_T(m_{H^{\pm}}^2,m_X^2) (c_\beta R^A_{22} -s_\beta R^A_{21})^2   \right. \nonumber \\
&&\left. -3 F_T(m_{Z}^2,m_{ref}^2) + 3F_T(m_{W}^2,m_{ref}^2)\right], 
\eea
where
\bea\label{stu-ft}
F_T(a,b)&=&\frac{1}{2}(a+b)-\frac{ab}{a-b}\log(\frac{a}{b}),~~F_S(a,b,c)=B_{22}(a,b,c)-B_{22}(0,b,c),\nonumber\\
F_{S0}(a,b,c)&=&B_{0}(a,b,c)-B_{0}(0,b,c)
\eea
with
\bea
&&B_{22}(a,b,c)=\frac{1}{4}\left[b+c-\frac{1}{3}a\right] - \frac{1}{2}\int^1_0 dx~X\log(X-i\epsilon),\nonumber\\
&&B_{0}(a,b,c) = -\int^1_0 dx ~ \log{(X-i\epsilon)}\, ,\nonumber\\
&&
X=bx+c(1-x)-ax(1-x).
\eea

\section{Excesses around 95.4 GeV} 
The assumed origins of the diphoton and $b\bar{b}$ excesses around the 95.4 GeV are the resonant
productions of the lightest CP-even Higgs $h_1$ in the model.
In the narrow width approximation, these signal strengths can be expressed as follows:
\begin{align}
	\mu_{b\bar{b}}&=\frac{\sigma_\text{2HDMS}(e^+ e^-\rightarrow Z h_1)}{\sigma_\text{SM}(e^+ e^-\rightarrow Z h_{95.4}^\text{SM})}\times\frac{\text{BR}_\text{2HDMS}(h_1\rightarrow b\bar{b})}{\text{BR}_\text{SM}(h_{95.4}^\text{SM}\rightarrow b\bar{b})}=|y^{h_1}_V|^2\frac{\text{BR}_\text{2HDMS}(h_1\rightarrow b\bar{b})}{\text{BR}_\text{SM}(h_{95.4}^\text{SM}\rightarrow b\bar{b})}
	\label{eq:mulep}
	\end{align}
	\begin{align}
	\mu_{\gamma\gamma}&=\frac{\sigma_\text{2HDMS}(g g\rightarrow h_1)}{\sigma_\text{SM}(g g\rightarrow h_{95.4}^\text{SM})}\times \frac{\text{BR}_\text{2HDMS}(h_1\rightarrow \gamma\gamma)}{\text{BR}_\text{SM}(h_{95.4}^\text{SM}\rightarrow \gamma\gamma)}\simeq |y^{h_1}_u|^2\frac{\text{BR}_\text{2HDMS}(h_1\rightarrow \gamma\gamma)}{\text{BR}_\text{SM}(h_{95.4}^\text{SM}\rightarrow \gamma\gamma)}
	\label{eq:mucms}
\end{align}

In our calculations, we take the mixing parameters and Yukawa parameters as follows: 
\begin{align}\label{parameter}
 & 0 \leq c_2 \leq 0.4,~~~ s_{\beta 13} = 1.0,~~~ 0.5 \leq \tan\beta \leq 2,\nonumber\\
 & 0 \leq s_{\beta 1} \leq 1.0,~~~ 0.2 \leq s_4 \leq 0.7,\nonumber\\
 & \kappa_u=0,~~~\kappa_\ell=0,~~~ -5.0 \leq \kappa_d \leq 5.0.
\end{align}
For $s_{\beta 13} = 1.0$ and $c_2 \to 0$, the couplings between the $h_2$ and SM particles approach to those of SM, which is favored by the signal data of 
the 125 GeV Higgs. In order to weaken the bounds on the extra Higgses from the direct searches at the LHC and indirect searches via the flavour
physics data, we choose $\kappa_u=\kappa_\ell=0$ and $-5.0 \leq \kappa_d \leq 5.0$. The nonzero $\kappa_d$ can make $y_d^{h_1}$ to be smaller than $y_V^{h_1}$
and $y_u^{h_1}$, and suppress the total width of $h_1$ significantly. As a result, $Br(h_1\to \gamma\gamma)$ is sizably enhanced.

\begin{figure}[tb]
\centering
\includegraphics[width=16.cm]{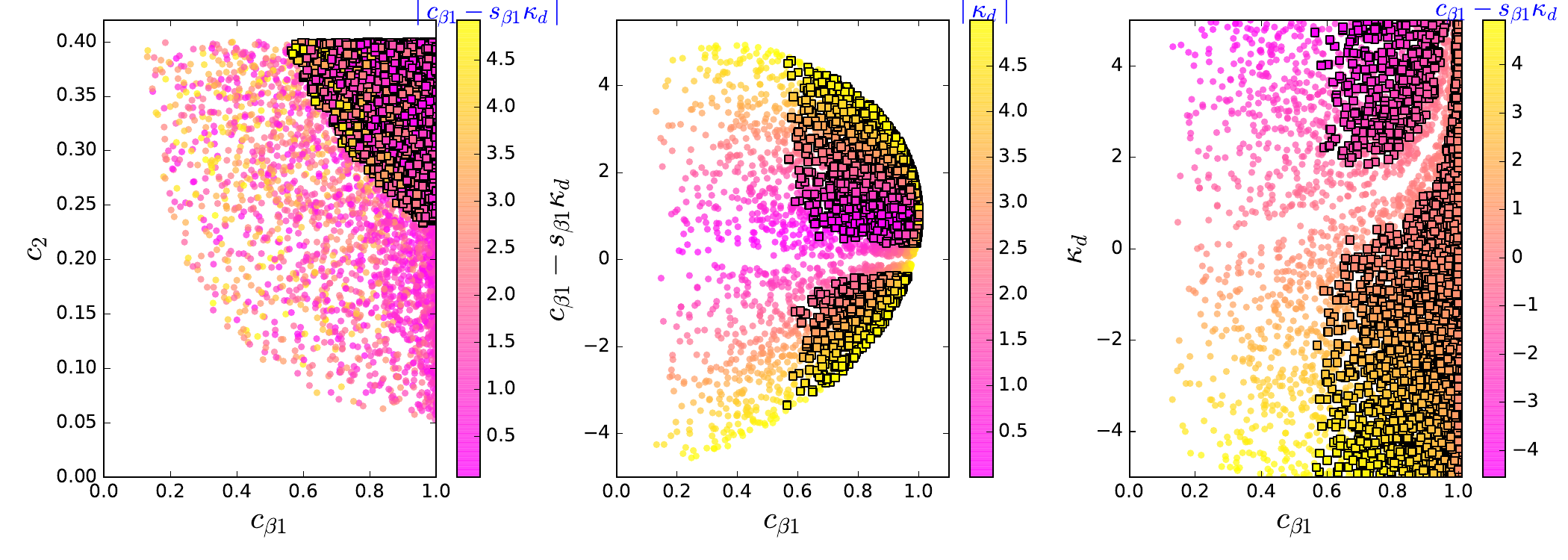}
\vspace{-0.4cm}\caption{The samples accommodating the $b\bar{b}$ excess at the 95.4 GeV. The excess is explained
 within $1\sigma$ and $2\sigma$ ranges for the squares and circles, respectively.}
\label{figbb}
\end{figure}

\begin{figure}[tb]
\centering
\includegraphics[width=16.cm]{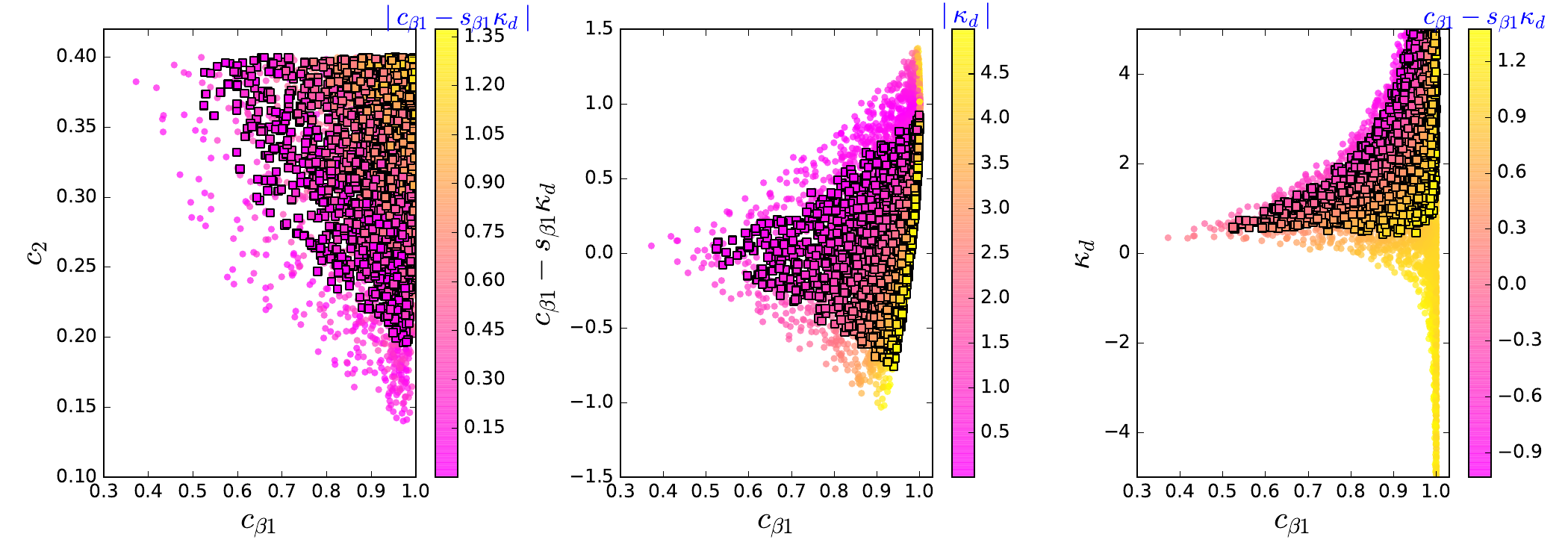}
\vspace{-0.4cm}\caption{Same as the Fig. \ref{figbb}, but for the diphoton excess at the 95.4 GeV.}
\label{figrr}
\end{figure}

\begin{figure}[tb]
\centering
\includegraphics[width=12.cm]{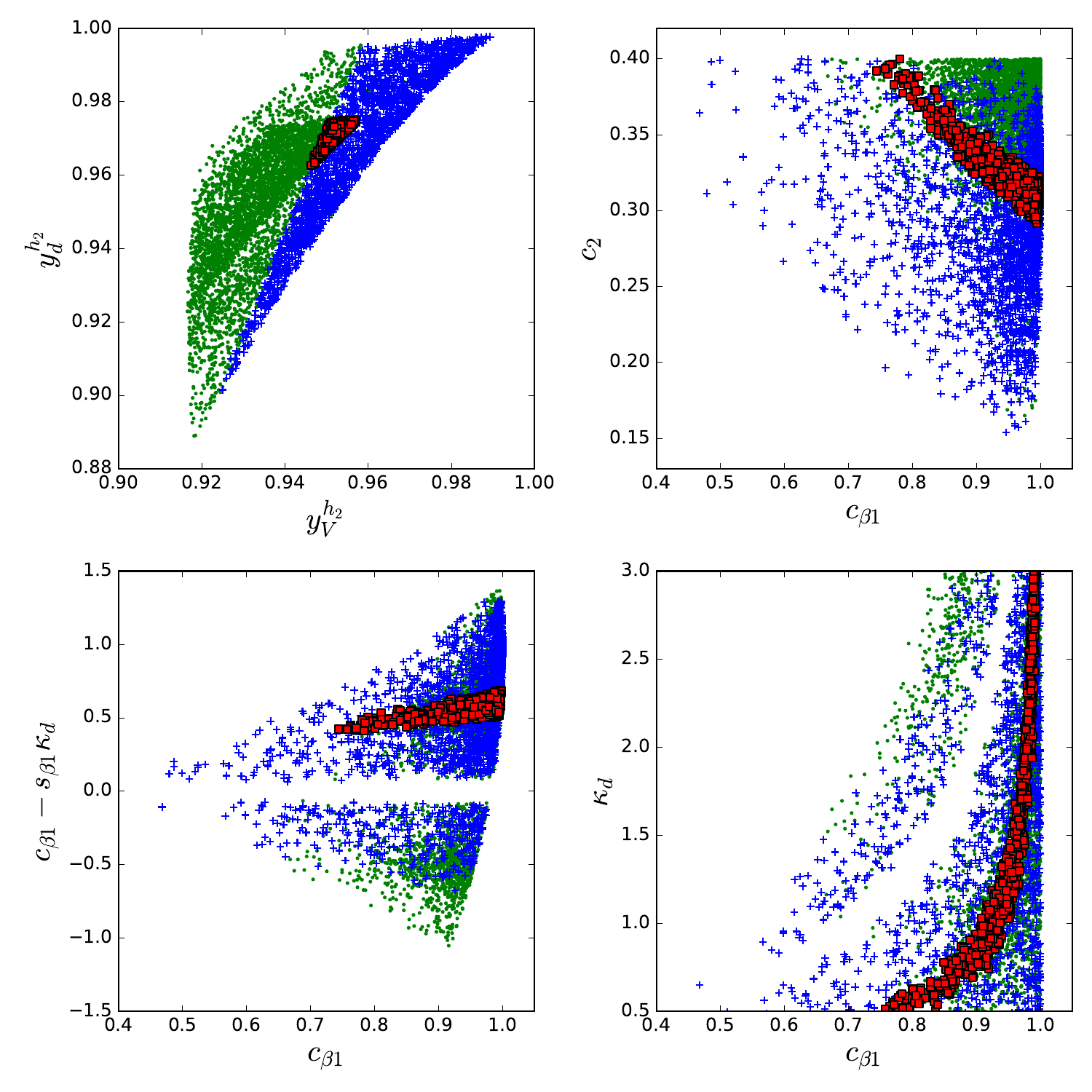}
\vspace{-0.4cm}\caption{The circles (green) are excluded by the constraints of "pre-95" while the pluses (blue) and squares (red) are allowed.
The $b\bar{b}$ and diphoton excesses are simultaneously explained
 within $1\sigma$ and $2\sigma$ ranges for squares (red) and the pluses (blue).}
\label{figbbrr}
\end{figure}
 
In Fig. \ref{figbb} and Fig. \ref{figrr} we respectively display the parameter points explaining the $b\bar{b}$ and diphoton excesses at the 95.4 GeV
 within $1\sigma$ and $2\sigma$ ranges without imposing any other constraints. From  Fig. \ref{figbb}, it can be observed that
 $c_2$ is imposed a lower bound for a given $c_{\beta 1}$ since the signal strength $\mu_{b\bar{b}}$ is 
directly proportional to $|y_{h_1}^V|^2$. The Yukawa coupling of bottom quark to $h_1$ is proportional to 
$c_{\beta 1} -s_{\beta 1} \kappa_d$, and therefore a relative large value of $|c_{\beta 1} -s_{\beta 1} \kappa_d|$
is favored to explain the $b\bar{b}$ excess, especially for the explanation within $1 \sigma$ range.
The allowed $\kappa_d$ and $c_{\beta 1}$ are displayed in the right panel of Fig. \ref{figbb}.
Different from the case of the $b\bar{b}$ excess, the explanation of diphoton excess at the 95.4 GeV favors
a relative small $|c_{\beta 1} -s_{\beta 1} \kappa_d|$, which can suppress the total width of $h_1$, and then
enhance $Br(h_1\to \gamma\gamma)$. As a result, a positive $\kappa_d$ is favored for the case of $c_{\beta 1} >0$ and $s_{\beta 1} >0$.
When $c_{\beta 1}$ approaches to 1.0, a negative $\kappa_d$ is allowed to explain the diphoton excess with $2\sigma$ range, as shown
in the right panel of Fig. \ref{figrr}.

After imposing the constraints of "pre-95" (denoting theoretical constraints, the
oblique parameters, the signal data of the 125 GeV Higgs, the searches for extra Higgses at the LHC and the flavor observables),
we show the parameter points explaining the diphoton and $b\bar{b}$ excesses at the 95.4 GeV in Fig. \ref{figbbrr}.
We find that the whole parameter space chosen in the Eq. (\ref{parameter}) can be consistent with the measurement of $B\to X_s\gamma$.
The choice of $\kappa_u=0$ can suppress the top Yukawa couplings to $H,~A,~X$, and $H^\pm$, and 
reduce the production cross sections of these extra Higgses at the LHC. As a result, the exclusion bounds of searches for additional Higgs bosons at the LHC
can be markedly weakened. The signal data of the 125 GeV Higgs impose stringent constraints on $y^{h_2}_V$ and $y^{h_2}_d$, and
exclude many parameter points explaining the $b\bar{b}$ and diphoton excesses. 
In the region of $0.15<c_2<0.4$ and $0.45<c_{\beta 1}<1.0$, the model can simultaneously explain the $b\bar{b}$ and diphoton excesses at the 95.4 GeV within $2\sigma$ ranges.
The $|\kappa_d|$ is favored to decrease with $c_{\beta 1}$, and $0<\kappa_d<1$ is required for the case of $c_{\beta 1}<0.6$. 
There is a small region explaining both $b\bar{b}$ and diphoton excesses within $1\sigma$ range, in which $c_2$, $c_{\beta 1}$, and $\kappa_d$ have strong dependencies.

\section{Phase transition}
To examine the PTs we first study the effective scalar potential of the model at finite
temperature. The neutral elements of $\Phi_1$ and $\Phi_2$ are parametrized as $\frac{\varphi_1}{\sqrt{2}}$ and $\frac{\varphi_2+i\varphi_3}{\sqrt{2}}$, and the singlet
scalar field is parametrized as $\frac{\chi+i\eta_s}{\sqrt{2}}$.
It is plausible to choose the imaginary part of the neutral component of $\Phi_1$ to be zero because the potential of Eq. (\ref{V2HDM}) only depends on the relative phase of the
neutral components of $\Phi_1$ and $\Phi_2$ .

The full effective potential at finite temperature is gauge-dependent \cite{vgauge1,vgauge2}, which is composed of four parts: the tree level potential, the
Coleman-Weinberg term \cite{vcw}, the finite temperature corrections \cite{vloop} and the resummed daisy corrections \cite{vring1,vring2}. 
In this paper we consider the high-temperature approximation of effective potential, which keeps only the tree-level potential
and the thermal mass terms in the high-temperature expansion. Therefore, the effective potential is gauge invariant, and it does not depend on the
resummation scheme and the renormalization scheme. 
The high-temperature approximation of effective potential is written as 
\begin{align} \label{veff0}
&V_{eff}(\varphi_1,\varphi_2,\varphi_3,\chi,\eta_s) =\frac{1}{2} (m_{11}^2 + \Pi_{\varphi_1}) \varphi_1^2+ \frac{1}{2} (m_{22}^2+ \Pi_{\varphi_2}) (\varphi_2^2 + \varphi_3^2)+\frac{1}{2} (m_{S}^2 + {m^\prime_S}^{2} +\Pi_{\chi}) \chi^2
\nonumber\\
&+ \frac{1}{2} (m_{S}^2-{{m^\prime_S}^{2}} +\Pi_{\eta_s}) \eta_s^2 
+ \frac{1}{8}(\lambda_1\varphi_1^4+\lambda_2\varphi_2^4 +\lambda_2 \varphi_3^4)+\frac{1}{4} \lambda_{345}\varphi_1^2\varphi_2^2 +\frac{1}{4} \bar{\lambda}_{345} \varphi_1^2 \varphi_3^2\nonumber\\
&+\frac{\lambda_2}{4}\varphi_3^2\varphi_2^2-m_{12}^2\varphi_1\varphi_2-\frac{\mu}{\sqrt{2}}\varphi_3\eta_s\varphi_1
+\frac{\lambda^{\prime}_1}{4}(\chi^2+\eta_s^2)\varphi_1^2+\frac{\lambda^{\prime}_2}{4}(\chi^2+\eta_s^2)(\varphi_2^2+\varphi_3^2) \nonumber\\
&+ \frac{\lambda^{\prime}_4}{2}(\chi^2-\eta_s^2)\varphi_1^2 +\frac{\lambda^{\prime}_5}{2}(\chi^2-\eta_s^2)( \varphi_3^2+\varphi_2^2)
+ (\frac{\lambda^{''}_1}{48}+\frac{\lambda^{\prime\prime}_3}{16}) (\chi^4+\eta_s^4)+\frac{1}{8}(\lambda^{\prime\prime}_3-\lambda^{\prime\prime}_1)\chi^2\eta_s^2\nonumber\\
&+\frac{\lambda^{''}_2}{12}(\chi^4-\eta_s^4)+\frac{\sqrt{2}}{3}\mu_{s}\chi^3,\nonumber\\
&\Pi_{\varphi_1}= \left[{9g^2\over 2} + {3g'^2\over 2} + 6\lam_{1} +4\lam_{3} +2\lam_4 + 2\lambda^\prime_{1} + 6y_t^2 (c_\beta-s_\beta \kappa_u)^2
 + 6y_b^2 (c_\beta-s_\beta \kappa_d)^2\right] {T^2 \over 24},\nonumber\\
&\Pi_{\varphi_2}= \left[{9g^2\over 2} + {3g'^2\over 2} + 6\lam_{2} +4\lam_{3} +2\lam_4 + 2\lambda^\prime_{2} + 6y_t^2 (s_\beta+c_\beta \kappa_u)^2
 + 6y_b^2 (s_\beta+c_\beta \kappa_d)^2\right] {T^2 \over 24},\nonumber\\
&\Pi_{\varphi_3} =\Pi_{\varphi_2},\nonumber\\
&\Pi_{\chi}= \left[ 4\lambda^\prime_{1} + 4\lambda^\prime_{2} +2{\lambda^{\prime\prime}_2}+2{\lambda^{\prime\prime}_3}+8\lambda^\prime_{4} + 8\lambda^\prime_{5} \right] {T^2 \over 24},\nonumber\\
&\Pi_{\eta_s}=   \left[ 4\lambda^\prime_{1} + 4\lambda^\prime_{2} -2{\lambda^{\prime\prime}_2}+2{\lambda^{\prime\prime}_3}-8\lambda^\prime_{4} -8\lambda^\prime_{5} \right]  {T^2 \over 24},
\end{align}
where $\bar{\lambda}_{345}=\lambda_3+\lambda_4-\lambda_5$, $y_t={\sqrt{2} m_t \over v}$,
$y_b={\sqrt{2} m_b \over v}$, and $\Pi_{i}$ denotes the thermal mass terms of the field $i$.

In a first-order PT, the bubble nucleation rate per unit volume at the finite temperature is given by \cite{bubble-0,bubble-1,bubble-2}
\begin{eqnarray}
\Gamma \ \approx \ A(T)e^{-S_3/T},
\end{eqnarray}
where $A(T)\sim T^4$ is a prefactor and $S_3$ is a three-dimensional Euclidian action,
\begin{equation}
S_3 = 4 \pi \int_0^\infty \mathrm{d}r \, 
r^2 \left [
	\sum_{i=1}^5 \frac{ 1 }{ 2 } 
		\left( 
		\frac{ \mathrm{d} \phi_i }{ \mathrm{d} r }
		\right)^2
	+ V_{eff}
\right ].
\end{equation}
The $\phi_{i=1,2,3,4,5}$ denotes $\varphi_1$, $\varphi_2$, $\varphi_3$, $\chi$, and $\eta_s$, which are determined by bounce equations~\cite{vaccum-eq}
\begin{equation}
\label{eq: bubble_equation}
\frac{ \mathrm{d}^2 \phi_i }{ \mathrm{d} r^2 } + \frac{ 2 }{ r } \frac{ \mathrm{d} \phi_i }{ \mathrm{d} r } = \frac{ \partial V_{eff} }{ \partial \phi_i }, \quad ( i=1,2,3,4,5). 
\end{equation}
We employ $\textsf{FindBounce}$ to solve the bounce equations, and obtain the bubble wall VEV profiles \cite{findbounce}. 
At the nucleation temperature $T_n$, the thermal tunneling probability for bubble nucleation per horizon volume and per horizon time is 
of order one, and this condition is roughly converted to the relation $\frac{S_3(T)}{T}|_{T = T_n}\approx 140$. 

The CP symmetry is conserved both at very high temperature and the present temperature, and it is only spontaneously broken
at the finite temperature. Therefore, the Universe need undergo multi-step PTs.  It is noted that 
the effective potential $V_{eff}$ in Eq. (\ref{veff0}) has a $Z_2$ symmetry: $\varphi_3\to -\varphi_3$ and $\eta_s\to -\eta_s$. Therefore, there
will not be a bias between transitions to $(\langle \varphi_1 \rangle, \langle \varphi_2\rangle, \langle \varphi_3\rangle, \langle \chi \rangle, \langle \eta_s\rangle)$ 
and $(\langle \varphi_1 \rangle, \langle \varphi_2 \rangle, -\langle \varphi_3 \rangle, \langle \chi \rangle, -\langle \eta_s \rangle)$ from the origin (0,~0,~0,~0,~0) GeV.
The two kinds of bubbles will create
baryon asymmetry of opposite signs. As a result, the net BAU averaged over
the entire Universe will be zero resulting in a null BAU.
The issue can be solved in a special pattern of three-step PTs.
The first step is the $Z_2$-breaking PT, and the second step is a strongly first-order electroweak PT during which the BAU is generated.
Finally, through the third-step PT the observed vacuum is obtained and the CP symmetry is recovered at the present temperature.
During the first $Z_2$-breaking PT, different patches of the Universe would end up in either +$\langle \eta_s \rangle$ or -$\langle \eta_s \rangle$
vacuum, forming a network of domains. 
One may add a $Z_2$ symmetry breaking term, $-i\mu_3(S-S^*)^3$, which leads to a potential
difference between the vacua with $\pm\langle \eta_s\rangle$, $\Delta V$. 
The amount of bias $\Delta V$ needed for the deeper minimum
to be the only vacuum remaining when the next electroweak PT transition happens is very tiny,
 $\Delta V/T^4>10^{-16}$ \cite{1110.2876,bgs-1}.
Despite the $\mu_3$ term breaks the CP symmetry explicitly, the CP-conservation is still a safe approximation in the model because of a negligible value of $\mu_3$.

\begin{table}[t]
\centering
\begin{tabular}{| c | c | c |c | c | c | c |c | c | c | c |c | c | c | c | c | c | c |c | c | c | c |c |c |}
\hline
& $m_{h_1}$(GeV)& $m_{h_2}$(GeV) &$m_{h_3}=m_{H^\pm}$(GeV) & $m_{A}$(GeV) & $m_{X}$(GeV)& $m_{12}^2$(GeV$)^2$& $m^{\prime 2}_s$(GeV)$^2$ \\
\hline
BP1  & 95.4  &  125.0  & 451.42 &  80.99  & 412.39 & 5314.36 & 4720.09\\
 \hline
BP2  & 95.4  &  125.0  & 570.22   &  70.40  & 275.10 & 2482.07 & 5067.11 \\
 \hline
\end{tabular}
\centering
\begin{tabular}{| c | c | c| c | c | c |c | c | c | c |c | c | c | c |c | c | c | c | c | c | c |c | c | c | c |c |c | c |c |}
\hline
& $t_\beta$& $s_{\beta13}$ &$c_{2}$ & $s_{\beta1}$  & $s_{4}$  & $\lambda^{\prime}_5$  & $\lambda^{\prime\prime}_1$  & $\lambda^{\prime\prime}_2$ & $\kappa_d$ &$\kappa_u=\kappa_\ell$ & $\mu_s$(GeV) & $v_s$(GeV)&$\mu_{b\bar{b}}$ &$\mu_{\gamma\gamma}$\\
\hline
BP1& 1.147 &1.0 & 0.277 & 0.382 & 0.317& -1.429 & 3.908 & 3.166 & 0.954 & 0.0 &73.48  & -50.11 &0.0489   & 0.118  \\
 \hline
BP2&0.950 &1.0 & 0.247 & 0.498 & 0.426 & -1.024 & 4.384 & 2.324 & 0.891 & 0.0 &101.63  & -67.05&0.0280    &0.104   \\
 \hline
\end{tabular}
\caption{Input parameters for the BP1 and the BP2.}
\label{tabbp1}
\end{table}

\begin{figure}[tb]
\centering
\includegraphics[width=12.cm]{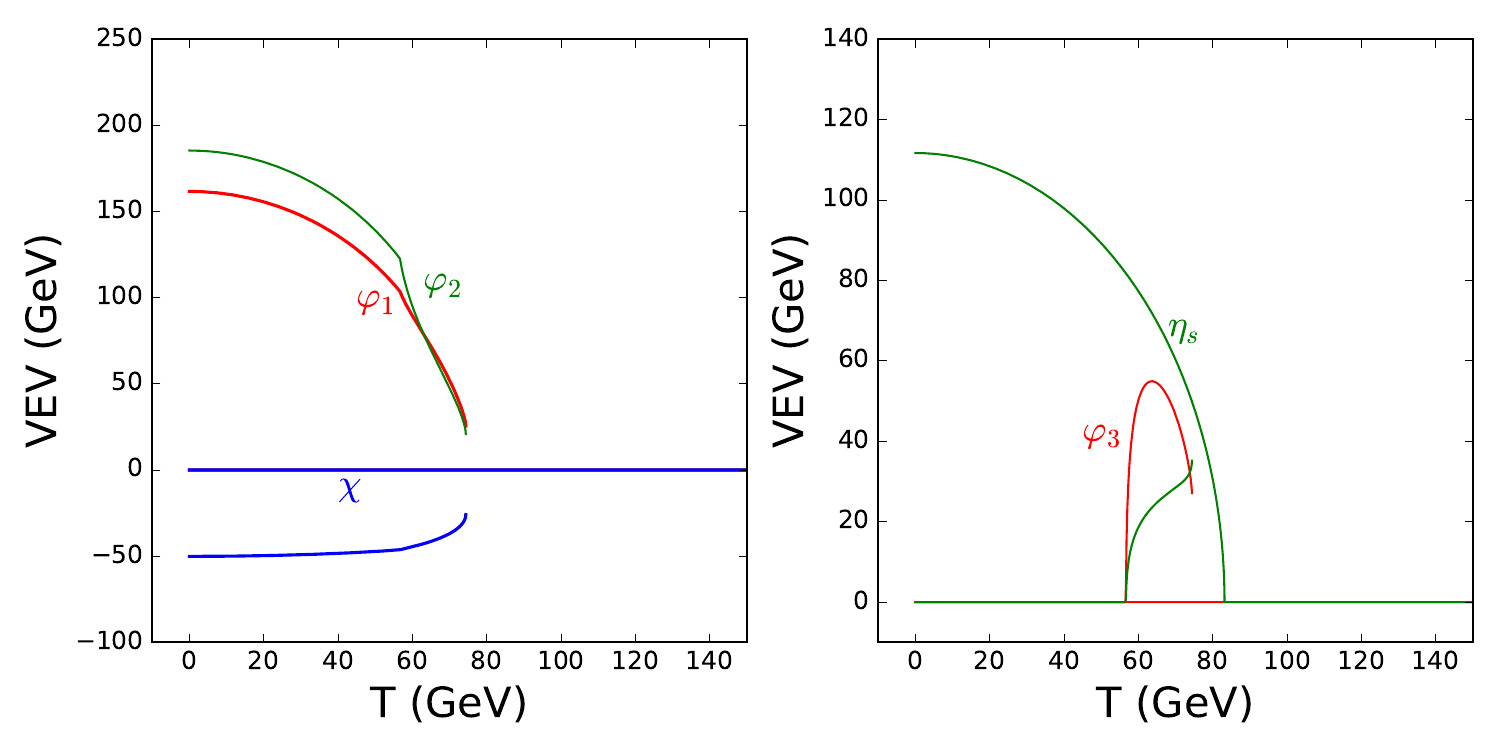}
\vspace{-0.4cm}\caption{Phase histories for the BP1. $\phi(i)$ denotes the VEV
of $\phi$ at the i-th phase, where $\phi=\varphi_1,\varphi_2,\chi,\varphi_3,\eta_s$. The black dot represents the occurrence of a second-order PT.}
\label{figbp1}
\end{figure}

\begin{figure}[tb]
\centering
\includegraphics[width=12.cm]{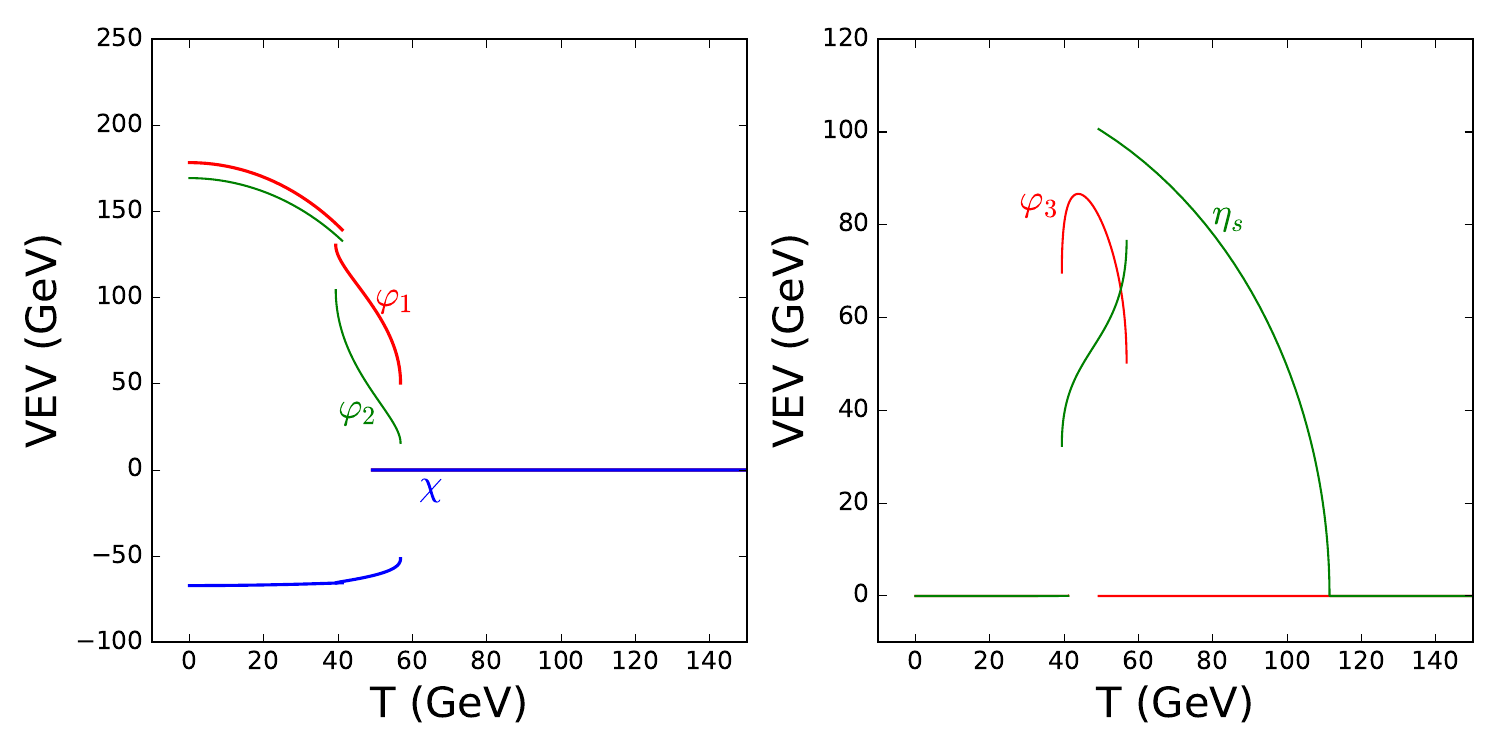}
\vspace{-0.4cm}\caption{Same as Fig. 1, but for the BP2.}
\label{figbp2}
\end{figure}

\begin{figure}[tb]
\centering
\includegraphics[width=16.cm]{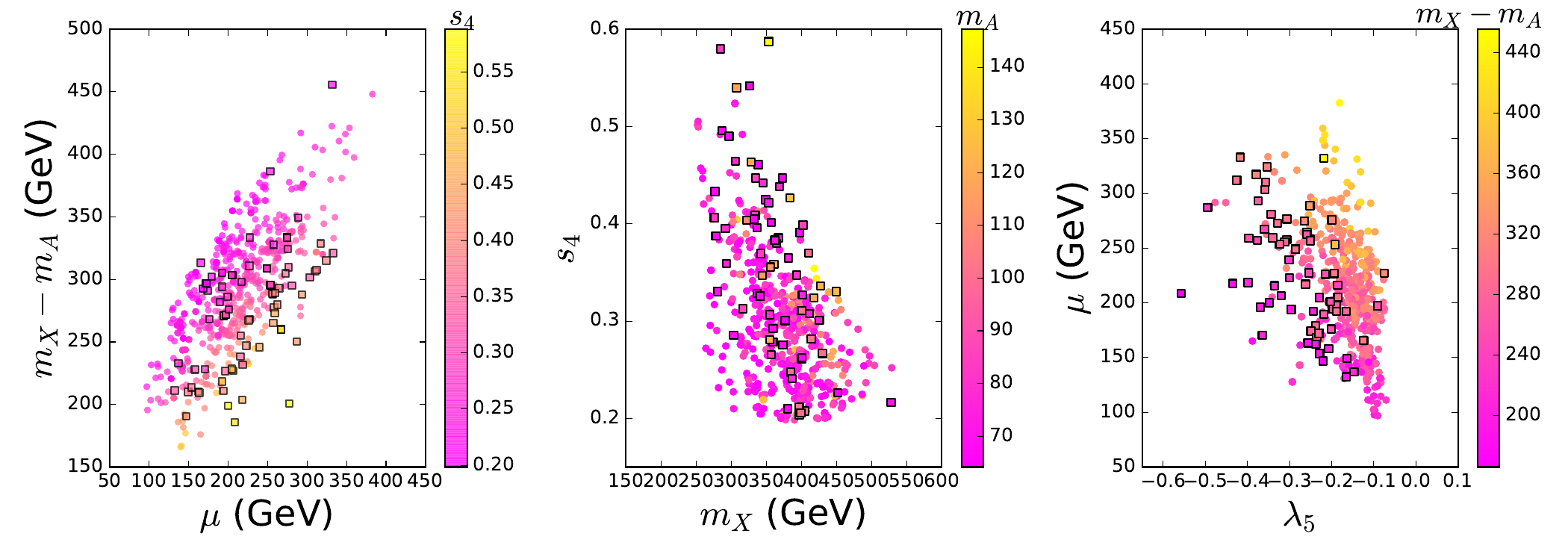}
\vspace{-0.4cm}\caption{The scattering plots achieving the three-step PTs with the characteristics mentioned in
the text. The third step is respectively first-order PT and second-order PT for the squares and circles.}
\label{figscan}
\end{figure}

We apply $\textsf{CosmoTransitions}$ to analyze the PTs \cite{cosmopt}.
After imposing the constraints of "pre-95" and requirement of explaining both $b\bar{b}$ and diphoton excesses within $2\sigma$ ranges, 
we find two different patterns of the three-step PTs with the characteristics mentioned above, 
and take two benchmark points BP1 and BP2 to discuss the PTs detailedly. The key input parameters are
displayed in Table \ref{tabbp1}.  Fig. \ref{figbp1} and Fig. \ref{figbp2} exhibit the phase histories for the BP1 and BP2.
For the BP1, at a very high temperature, the minimum of the potential is at the origin 
since the contributions of the thermal mass terms to the effective potential are proportional to $T^2$.
When the temperature decreases to T=83.15 GeV, a second-order PT starts during which 
$\eta_s$ develops a nonzero VEV and the other four fields still remain zero.
At T=65.81 GeV, a strongly first-order electroweak PT happens which breaks electroweak symmetry,  (0, 0, 0, 0, 68.27) GeV $\to$ (68.88, 66.0, 53.86, -40.83, 25.47) GeV
 for ($\langle \varphi_1\rangle$, $\langle \varphi_2\rangle$, $\langle \varphi_3\rangle$, $\langle\chi\rangle$, $\langle\eta_s\rangle$). 
The strength of PT ($\xi=\frac{\sqrt{\varphi_1^2+\varphi_2^2+\varphi_3^2}}{T})$ is 1.66, and the BAU is generated via the EWBG mechanism.
At T=56.59 GeV, another second-order PT takes place. Next, the vacuum evolves along the final phase, and
 ultimately ends in the observed values at T = 0 GeV while the CP symmetry is restored. Meanwhile, $\xi>1$ is always remained so that the BAU is not washed out by
the sphaleron process \cite{pt-stren}.

For the BP2, the first step is second-order PT and the second step is strongly first-order PT, which are similar to those of BP1.
However, the third step is first-order PT during which the observed vacuum is obtained and the CP symmetry is restored.

There are no explicit CP-violating interactions in the scalar potential in Eq. (\ref{V2HDM}). 
Due to the thermal corrections to the effective potential in Eq. (\ref{veff0}), the $\eta_s$ field firstly acquires a nonzero VEV at finite temperature. 
Naturally, the mass term and interaction terms of the $S$ field will play an important role. 
In addition, the $\mu$ term (see Eq. (\ref{V2HDM}) and Eq. (\ref{veff0})) can lead to a close correlation between 
$\langle \varphi_3\rangle$ and $\langle \eta_s \rangle$ of the potential minimum, which plays a key role in the $\varphi_3$ field 
acquiring a nonzero VEV as the temperature evolves. Thus, an imaginary part of the top quark mass is induced by the interaction of $t\bar{t}(\varphi_2+i \varphi_3)$.

In Fig. \ref{figscan} we display some parameter points achieving the three-step PTs.
From Fig. \ref{figscan}, we find that the three-step PTs satisfying our requirements favor an appropriate value of $\mu$. 
As a result, according to Eq. (\ref{eq:lambdas}), $m_A$ and $m_X$ is required to have a large mass splitting, and the mass splitting
tends to increase with a decrease of $s_4$. Also $m_X$ is favored to have a large value for a small $s_4$. Besides,
the PT, whose third step is first order, tends to be achieved in the region of relatively large $\mid\lambda_5\mid$.

In BP1 and BP2, the 70-80 GeV lightest CP-odd scalar
boson $A$ is predicted. Because of $\kappa_u=\kappa_\ell=0$, $\kappa_d=0.954$ (0.891) for BP1 and BP2, 
the $A$ coupling to the top quark is absent according to Eq. (\ref{hffcoupling}), 
leading to a negligible production cross section via gluon-gluon fusion process at the LHC.
The $A$ can be mainly produced via the $pp\to Ab\bar{b}$ process at the LHC, and then $A$ decays into $b\bar{b}$.
 However, it is challenging to isolate potential signals from the overwhelming QCD background, and several strategies need be 
improved, such as precise event selection with optimized kinematic cuts, and advanced jet substructure techniques to distinguish signal-like jets from QCD jets.
 Additionally, the machine learning-based analyses can enhance signal-to-background discrimination.

\section{Baryogenesis}
We take the WKB approximation method to calculate the CP-violating source terms and chemical potentials transport equations of particle species in the wall frame with 
a radial coordinate $z$ \cite{bg2h-3,0006119,0604159}.
When a top quark penetrates the bubble wall, it acquires a complex mass as a function of $z$,
\beq
m_t(z)=\frac{y_t}{\sqrt{2}}\sqrt{(c_\beta \varphi_1(z)+s_\beta \varphi_2(z))^2+s_\beta^2 \varphi^2_3(z)}~ e^{i\Theta_t(z)}
\eeq
 with
\bea&&\Theta_t(z)=\Theta_Z(z)+\arctan\frac{s_\beta \varphi_3(z)}{c_\beta \varphi_1(z)+s_\beta \varphi_2(z)},
\partial_z\Theta_Z(z)=-\frac{\varphi^2_2(z)+\varphi^2_3(z)}{\varphi^2_1(z)+\varphi^2_2(z)+\varphi^2_3(z)}\partial_z\Theta(z),\nonumber\\
&&\Theta(z)=\arctan\frac{\varphi_3(z)}{\varphi_2(z)}.
\eea 
 Because the imaginary part of the neutral element of $\Phi_1$ is chosen to be zero in our discussions, 
 the nonvanishing $Z_\mu$ field produce an additional CP-violating force acting on the top quark, which is removed by a local axial transformation of top quark,
reintroducing an additional overall phase $\Theta_Z(z)$ into $m_t$ \cite{bg2h-4}.

The complex mass of the top quark drives the transport equations, which include effects of the strong sphaleron process ($\Gamma_{ss}$) \cite{bg2h-3,9311367}, 
W-scattering ($\Gamma_W$) \cite{bg2h-3,9506477}, the top Yukawa interaction ($\Gamma_y$) \cite{bg2h-3,9506477},
 the top helicity flips ($\Gamma_M$) \cite{bg2h-3,9506477}, and the Higgs number violation ($\Gamma_h$) \cite{bg2h-3,9506477}. The transport equations are given by
	\begin{align}
		0 =   & 3 v_W K_{1,t} \left( \partial_z \mu_{t,2} \right) + 3v_W K_{2,t} \left( \partial_z m_t^2 \right) \mu_{t,2} + 3 \left( \partial_z u_{t,2} \right) \notag
		\\ &- 3\Gamma_y \left(\mu_{t,2} + \mu_{t^c,2} + \mu_{h,2} \right) - 6\Gamma_M \left( \mu_{t,2} + \mu_{t^c,2} \right) - 3\Gamma_W \left( \mu_{t,2} - \mu_{b,2} \right) \notag
		\\ &- 3\Gamma_{ss} \left[ \left(1+9 K_{1,t} \right) \mu_{t,2} + \left(1+9 K_{1,b} \right) \mu_{b,2} + \left(1-9 K_{1,t} \right) \mu_{t^c,2} \right] \notag \,,\\
                0=    & 3 v_W K_{1,t} \left( \partial_z \mu_{t^c,2} \right)  + 3v_W K_{2,t} \left( \partial_z m_t^2 \right)  \mu_{t^c,2} + 3 \left( \partial_z u_{t^c,2} \right) \notag   \\
		      & - 3\Gamma_y \left(\mu_{t,2} + \mu_{b,2} + 2\mu_{t^c,2} + 2\mu_{h,2} \right) - 6\Gamma_M \left( \mu_{t,2} + \mu_{t^c,2} \right) \notag    \\
		      & - 3\Gamma_{ss} \left[ \left( 1+9 K_{1,t}\right) \mu_{t,2} + \left(1+9K_{1,b}\right) \mu_{b,2} + \left(1-9K_{1,t}\right) \mu_{t^c,2} \right]  \notag \,,          \\
		0 =   & 3v_W K_{1,b} \left(\partial_z \mu_{b,2}\right) + 3 \left(\partial_z u_{b,2} \right) - 3\Gamma_y \left( \mu_{b,2} + \mu_{t^c,2} + \mu_{h,2} \right) - 3\Gamma_W \left( \mu_{b,2} - \mu_{t,2} \right) \notag \,,   \\
		      & - 3\Gamma_{ss} \left[ \left( 1 + 9K_{1,t}\right) \mu_{t,2} + (1+9K_{1,b}) \mu_{b,2} + (1-9K_{1,t}) \mu_{t^c,2} \right] \notag \,, \\
		0 =   & 4v_W K_{1,h} \left( \partial_z \mu_{h,2}\right) +
		4\left( \partial_z u_{h,2}\right) - 3\Gamma_y \left(
		\mu_{t,2} + \mu_{b,2} + 2\mu_{t^c,2} + 2\mu_{h,2} \right) -
		4\Gamma_h
		\mu_{h,2}  \notag\,,\nonumber\\
		S_t = & -3K_{4,t} \left( \partial_z \mu_{t,2}\right) + 3v_W \tilde{K}_{5,t} \left( \partial_z u_{t,2}\right) + 3v_W \tilde{K}_{6,t} \left( \partial_z m_t^2 \right) u_{t,2} + 3\Gamma_t^\mathrm{tot} u_{t,2} \notag \,,       \\
		0 =   & -3K_{4,b} \left( \partial_z \mu_{b,2} \right) + 3v_W \tilde{K}_{5,b} \left(\partial_z u_{b,2}\right) + 3\Gamma_b^\mathrm{tot} u_{b,2}  \notag  \,,   \\
		S_t = & -3K_{4,t} \left( \partial_z \mu_{t^c,2}\right) + 3v_W \tilde{K}_{5,t} \left( \partial_z u_{t^c,2}\right) + 3v_W \tilde{K}_{6,t} \left( \partial_z m_t^2\right) u_{t^c,2} + 3\Gamma_t^\mathrm{tot} u_{t^c,2} \notag \,, \\
		0 =   & -4K_{4,h} \left( \partial_z \mu_{h,2} \right) + 4v_W \tilde{K}_{5,h} \left( \partial_z u_{h,2} \right) + 4\Gamma_h^\mathrm{tot} u_{h,2}  \,.
\label{TransportEquations}
\end{align}

The $\mu_{i,2}$ and $u_{i,2}$ are the second-order CP-odd chemical potential and the
plasma velocity of the particle $i=t,~t^c,~b,~h$. $S_t$ is the source term,
\beq \label{Eq:TransportEquations:Source}
	S_t = -v_W K_{8,t} \partial_z \left( m_t^2 \partial_z \Theta_t \right) + v_W K_{9,t} \left( \partial_z \Theta_t \right) m_t^2  \left( \partial_z m_t^2\right).
\eeq
The $\Gamma_{i}^{\mathrm{tot}}$ are the total reaction rate of the particle $i$ \cite{bg2h-3,0604159}, and 
the functions $K_{a,i}$ and $\tilde{K}_{a,i}$ ($a=1-9$) are given in Ref. \cite{0604159}.

We solve the transport equations with the boundary conditions $\mu_i$ $(z=\pm \infty) =0$ ($i=t,~t^c,~b,~h$), and obtain the chemical potentials $\mu_i$
and velocity perturbations $u_i$ of
each particle species. 
The weak sphaleron process converts the left-handed quark number into a baryon asymmetry, which is calculated as 
\begin{equation}
	Y_B = \frac{405 \Gamma_{ws}}{4\pi^2 v_w g_* T_{n}}\int_0^\infty dz \mu_{B_L}(z) f_{sph}(z) \exp\left(-\frac{45\Gamma_{ws}z}{4v_w}\right)\,.
\end{equation}
$\Gamma_{ws}\simeq 10^{-6} T_{n}$ is
 the weak sphaleron rate inside bubble \cite{sphaleron-ws} and $v_w$ is the wall velocity. 
$\mu_{B_L}$ is the chemical potential of left-handed quarks, and
the function $f_{sph}(z) = min(1, 2.4 \frac{T_n}{\Gamma_{ws}} e^{-40 v_n(z)/T_n} )$ with $v_n(z)=\sqrt{\langle \varphi_1(z)\rangle^2+\langle \varphi_2(z)\rangle^2+\langle \varphi_3(z)\rangle^2}$ 
is employed to smoothly interpolate between the sphaleron
rate in the unbroken and broken phases \cite{bg2h-4}.


\begin{figure}[tb]
\centering
\includegraphics[width=10.cm]{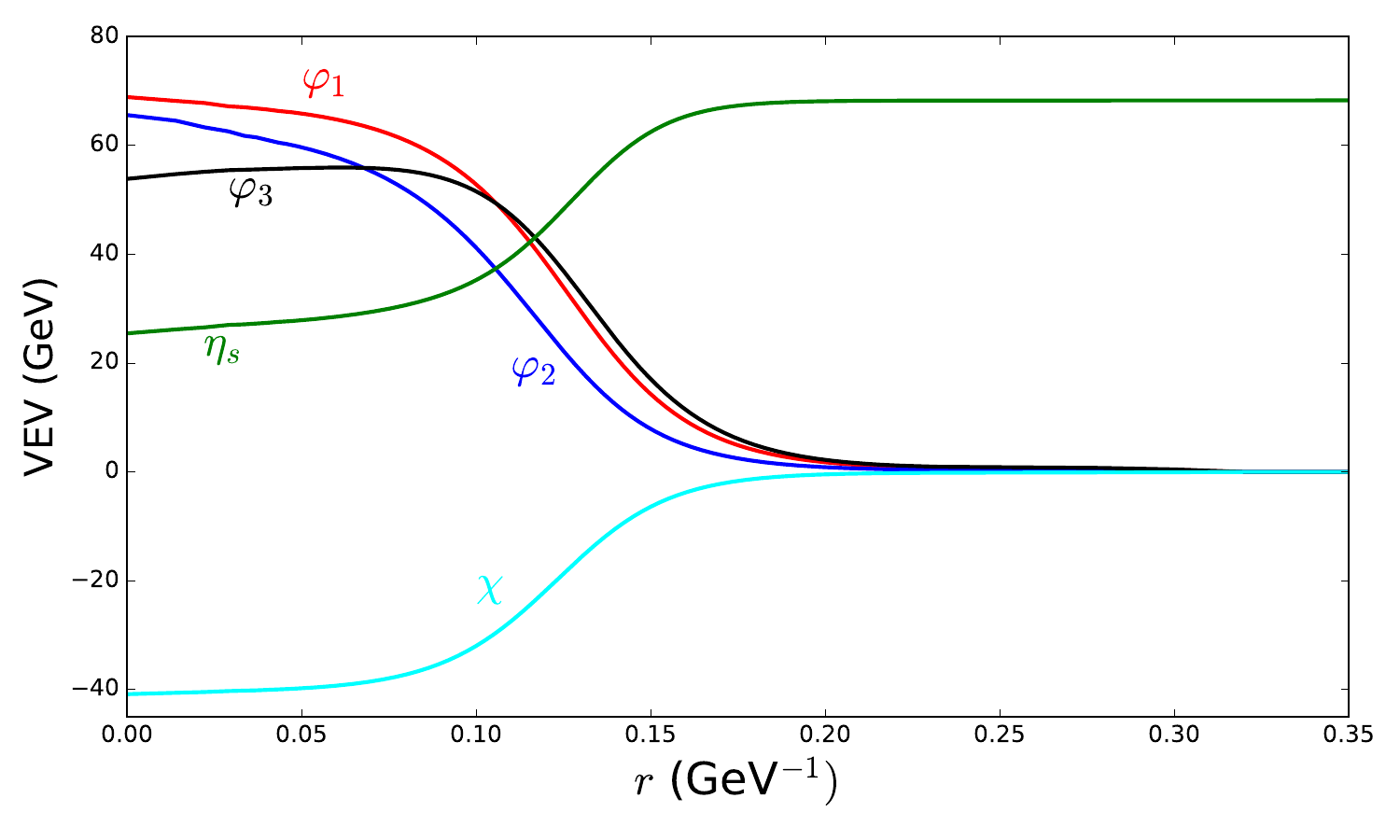}
\vspace{-0.4cm}\caption{The radial nucleation bubble wall VEV profiles for the second-step PT of the BP1.}
\label{figbounce}
\end{figure}

\begin{figure}[tb]
\centering
\includegraphics[width=6.5cm]{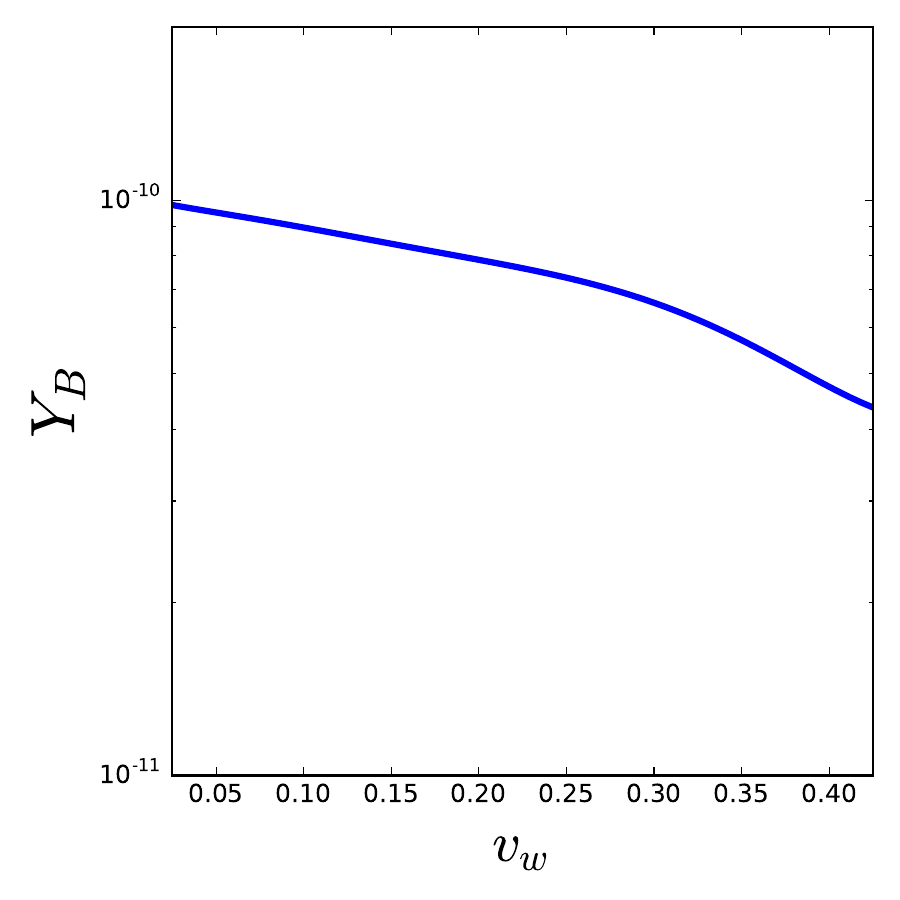}
\vspace{-0.8cm}\caption{$Y_B$ as a function of $v_w$ for the BP1.}
\label{figvw}
\end{figure}

The calculation of BAU for the BP1 depends on the bubble wall profiles for the second-step PT, and they are given 
in Fig. \ref{figbounce}. The WKB method of calculating transport equations needs the condition of $L_W T_n \gg 1$, where $L_W$ is the width of bubble wall.
The $L_W T_n$ for the BP1 is approximately estimated to be 2.6.
 Fig. \ref{figvw} displays $Y_B$ as a function of the wall velocity $v_w$ for the BP1.
$Y_B$ tends to decrease with an increase of $v_w$, and $8\times 10^{-11}<Y_B<10\times 10^{-11}$ for $0.025<v_W<0.19$.

\section{Conclusion}
In this work we studied a complex singlet scalar extension of the 2HDM in which the 95.4 GeV Higgs is from the mixing of three CP-even Higgs fields, and
there is a mixing between the two CP-odd Higgs fields.
Considering relevant theoretical and experimental constraints, we found that the model can simultaneously explain the diphoton and $b\bar{b}$ excesses at the 95.4 GeV
while achieving spontaneous CP-violation at the finite temperature. Some strong dependencies among several key parameters are displayed.
Finally, we picked out a benchmark point accommodating the diphoton and $b\bar{b}$ excesses simultaneously, and 
found that the model can explain BAU while satisfying the bound of EDM.

\section*{Acknowledgment}
This work was supported by the National Natural Science Foundation
of China under grant 11975013 and by the project ZR2024MA001 and ZR2023MA038 supported by Shandong Provincial Natural Science Foundation. 



\end{document}